\documentclass[letterpaper,twocolumn,10pt]{article}
\usepackage{usenix2019_v3}

\usepackage{nopageno}
\usepackage{amsmath}
\usepackage{amssymb}
\usepackage{booktabs}
\usepackage{graphicx}
\usepackage{subcaption}
\usepackage{setspace}
\usepackage{fancybox}
\usepackage{tikz}
\usetikzlibrary{arrows}
\usepackage[utf8]{inputenc}
\usepackage[english]{babel}

\newif\ifshowchanges
\showchangesfalse 

\ifshowchanges
\newcommand{\CHANGED}[1]{\textcolor{blue}{#1}}
\else 
\newcommand{\CHANGED}[1]{#1}
\fi 

\makeatletter
\renewcommand{\paragraph}{%
  \@startsection{paragraph}{4}%
  {\z@}{1ex \@plus 1ex \@minus .2ex}{-1em}%
  {\normalfont\normalsize\bfseries}%
}
\makeatother
\newtheorem{definition}{Definition}[section]

\newcommand{\nbc}[3]{
	{\colorbox{#3}{\bfseries\sffamily\scriptsize\textcolor{white}{#1}}}
	{\textcolor{#3}{\sf\small\textit{#2}}}}

\newif\ifsubmission
\submissiontrue

\ifsubmission
\newcommand{\mpx}[1]{}
\newcommand{\cjt}[1]{}
\newcommand{\cas}[1]{}
\newcommand{\todo}[1]{}
\else
\definecolor{mpcolor}{rgb}{0.1,0.9,0.1}
\newcommand{\mpx}[1]{\nbc{MP suggests:}{#1}{mpcolor}}

\definecolor{cjtcolor}{rgb}{0.1,0.1,0.9}
\newcommand{\cjt}[1]{\nbc{CJT:}{#1}{cjtcolor}}

\definecolor{cascolor}{rgb}{1,0.75,0}
\newcommand{\cas}[1]{\nbc{Cris suggests:}{#1}{cascolor}}

\definecolor{todocolor}{rgb}{1,0,0}
\newcommand{\todo}[1]{\nbc{TODO:}{#1}{todocolor}}
\fi

\newcommand{\code}[1]{\texttt{#1}}

\newenvironment{resultbox}
{	\noindent
	\begin{center}
		\begin{Sbox}
			\begin{minipage}{.9\linewidth}
				\smallskip
			}				
			{  				
			\end{minipage}
		\end{Sbox}\fbox{\TheSbox}
	\end{center}
}

\newcommand\nbVersions{5,386,239}

\newcommand\nbPackageCount{676,539}
\newcommand\nbEdgeCount{4,543,473}
\newcommand\nbMaintainerCount{199,327}
\newcommand\nbPackageReachTopMax{166,086}
\newcommand\nbPackageReachTopMin{134,774}
\newcommand\nbAvgPackageReachLatest{230}
\newcommand\nbCVEs{609}

\begin{document}

\date{}

\title{\Large \bf
Small World with High Risks:\\
A Study of Security Threats in the npm Ecosystem
}

\author{
\todo{Uncomment "{\textbackslash}submissiontrue" before submitting}
{\rm Markus Zimmermann}\\
Department of Computer Science\\
TU Darmstadt
\and
{\rm Cristian-Alexandru Staicu}\\
Department of Computer Science\\
TU Darmstadt
\and
{\rm Cam Tenny}\\
r2c
\and
{\rm Michael Pradel}\\
Department of Computer Science\\
TU Darmstadt
} 

\maketitle

\begin{abstract}
The popularity of JavaScript has lead to a large ecosystem of third-party packages available via the npm software package registry.
The open nature of npm has boosted its growth, providing over 800,000 free and reusable software packages.
Unfortunately, this open nature also causes security risks, as evidenced by recent incidents of single packages that broke or attacked software running on millions of computers.
This paper studies security risks for users of npm by systematically analyzing dependencies between packages, the maintainers responsible for these packages, and publicly reported security issues.
Studying the potential for running vulnerable or malicious code due to third-party dependencies, we find that individual packages could impact large parts of the entire ecosystem.
Moreover, a very small number of maintainer accounts could be used to inject malicious code into the majority of all packages, a problem that has been increasing over time.
Studying the potential for accidentally using vulnerable code, we find that lack of maintenance causes many packages to depend on vulnerable code, even years after a vulnerability has become public.
Our results provide evidence that npm suffers from single points of failure and that unmaintained packages threaten large code bases.
We discuss several mitigation techniques, such as trusted maintainers and total first-party security, and analyze their potential effectiveness.
\end{abstract}

\section{Introduction}
JavaScript has become one of the most widely used programming languages.
To support JavaScript developers with third-party code, the \emph{node package manager}, or short \emph{npm}, provides hundreds of thousands of free and reusable code packages.
The npm platform consists of an online database for searching packages suitable for given tasks and a package manager, which resolves and automatically installs dependencies.
Since its inception in 2010, npm has steadily grown into a collection of over 800,000 packages, as of February 2019, and will likely grow beyond this number.
As the primary source of third-party JavaScript packages for the client-side, server-side, and other platforms, npm is the centerpiece of a large and important software ecosystem.

The npm ecosystem is open by design, allowing arbitrary users to freely share and reuse code.
Reusing a package is as simple as invoking a single command, which will download and install the package and all its transitive dependencies.
Sharing a package with the community is similarly easy, making code available to all others without any restrictions or checks.
The openness of npm has enabled its growth, providing packages for any 
situation imaginable, ranging from small utility packages to complex web server 
frameworks and user interface libraries.

Perhaps unsurprisingly, npm's openness comes with security risks, as evidenced by several recent incidents that broke or attacked software running on millions of computers.
In March 2016, the removal of a small utility package called \emph{left-pad} caused a large percentage of all packages to become unavailable because they directly or indirectly depended on left-pad.\footnote{\url{https://www.infoworld.com/article/3047177/javascript/how-one-yanked-javascript-package-wreaked-havoc.html}}
In July 2018, compromising the credentials of the maintainer of the popular \emph{eslint-scope} package enabled an attacker to release a malicious version of the package, which tried to send local files to a remote server.\footnote{\url{https://github.com/eslint/eslint-scope/issues/39}}

Are these incidents unfortunate individual cases or first evidence of a more general problem?
Given the popularity of npm, better understanding its weak points is an important step toward securing this software ecosystem.
In this paper, we systematically study security risks in the npm ecosystem by analyzing package dependencies, maintainers of packages, and publicly reported security issues.
In particular, we study the potential of individual packages and maintainers to impact the security of large parts of the ecosystem, as well as the ability of the ecosystem to handle security issues.
Our analysis is based on a set of metrics defined on the package dependency graph and its evolution over time.
Overall, our study involves \nbVersions\ versions of packages, 
\nbMaintainerCount\ maintainers, and \nbCVEs\ publicly known security issues.

The overall finding is that the densely connected nature of the npm ecosystem introduces several weak spots.
Specifically, our results include:
\begin{itemize}
	\item Installing an average npm package introduces an implicit trust on 
	79 third-party packages and 39 maintainers, creating a 
	surprisingly large attack surface.
	
	\item Highly popular packages directly or indirectly influence many other packages (often more than 100,000) and are thus potential targets for injecting malware.
	
	\item Some maintainers have an impact on hundreds of thousands of packages. As a result, a very small number of compromised maintainer accounts suffices to inject malware into the majority of all packages.
	
	\item The influence of individual packages and maintainers has been continuously growing over the past few years, aggravating the risk of malware injection attacks.
	
	\item A significant percentage (up to 40\%) of all packages depend on code 
	with at least one publicly known vulnerability.
	
\end{itemize}

Overall, these findings are a call-to-arms for mitigating security risks on the npm ecosystem.
As a first step, we discuss several mitigation strategies and analyze their 
potential effectiveness.
One strategy would be a vetting process that yields trusted maintainers.
We show that about 140 of such maintainers (out of a total of more than 150,000) could halve the risk imposed by compromised maintainers.
Another strategy we discuss is to vet the code of new releases of certain 
packages. We show that this strategy reduces the security risk slightly slower 
than trusting  the involved maintainers, but it still scales reasonably well, 
i.e., trusting the top 300 packages reduces the risk by half. If a given 
package passes the vetting process for maintainers and code, we say it has ``perfect first-party security''. If all its transitive dependencies pass the vetting processes we say 
that it has ``perfect third-party security''. If both conditions are met, we consider it a ``fully secured package''. While achieving this 
property for all the packages in the ecosystem is infeasible, packages that are 
very often downloaded or that have several dependents should aim to achieve it. 

\section{Security Risks in the npm Ecosystem}

To set the stage for our study, we describe some security-relevant particularities of the npm ecosystem and introduce several threat models.

\subsection{Particularities of npm}
\label{sec:part}
\paragraph{Locked Dependencies}
In npm, dependencies are declared in a configuration file called package.json, which specifies the name of the dependent package and a version constraint. The version constraint either gives a specific version, i.e., the 
dependency is \emph{locked}, or specifies a range of compatible 
versions, e.g., newer than version~X. Each time an npm package is installed, all 
its dependencies are resolved to a specific version, which is automatically downloaded and installed. 

Therefore, the same package installed on two different machines or at two 
different times may download different versions of a dependency. To solve this 
problem, npm introduced package-lock.json, which developers can use to lock 
their transitive dependencies to a specific version until a new lock file is 
generated. That is, each package in the dependency tree is locked to a specific 
version.
In this way, users ensure uniform installation of their packages and coarse 
grained update of their dependencies. However, a major shortcoming of this 
approach is that if a vulnerability is fixed for a given dependency, the 
patched version is not installed until the package-lock.json file is 
regenerated.
In other words, developers have a choice between uniform distribution of their 
code and up-to-date dependencies.  Often they choose the later, which leads to 
a technical lag~\cite{DBLP:conf/icsm/DecanMC18} between the latest available 
version of a package and the one used by dependents.

\paragraph{Heavy Reuse}

Recent work~\cite{DBLP:conf/msr/KikasGDP17,DBLP:conf/wcre/DecanMC17} provides preliminary evidence that code reuse in npm differs significantly from other ecosystems. One of the main characteristic of the npm ecosystem is the high number of transitive dependencies. 
For example, when using the core of the popular Spring 
web framework in Java, a developer transitively depends on ten other packages. 
In contrast, the Express.js web framework transitively depends on 47 other packages. 

\paragraph{Micropackages}
Related to the reuse culture, another interesting characteristic of npm is the heavy reliance on packages that consist of only few lines of source code, which we call \emph{micropackages}. 
Related work documents this trend and warns about its 
dangers~\cite{Abdalkareem2017, DBLP:journals/corr/abs-1709-04638}. These 
packages are an important part of the ecosystem, yet they increase the  
surface for certain attacks as much as functionality heavy packages. 
This excessive fragmentation 
of the npm codebase can thus lead to very high number of dependencies. 

\paragraph{No Privilege Separation}
In contrast to, e.g., the Java security model in which a 
SecurityManager\footnote{\url{https://docs.oracle.com/javase/6/docs/api/java/lang/SecurityManager.html}}
 can restrict the access to sensitive APIs, JavaScript does not provide any 
 kind of privilege separation between code loaded from different packages.
That is, any third-party package has the full privileges of the entire application.
This situation is compounded by the fact that many npm packages run outside of a browser, in particular on the Node.js platform, which does not provide any kind of sandbox.
Instead, any third-party package can access, e.g., the file system and the network.

\paragraph{No Systematic Vetting}
The process of discovering vulnerabilities in npm packages is still in its 
infancy.
There currently is no systematic vetting process for code published on npm.
Instead, known vulnerabilities are mostly reported by individuals, who find 
them through manual analysis or in recent research work, e.g., injection 
vulnerabilities~\cite{DBLP:conf/ndss/StaicuPL18}, regular expression denial of 
service~\cite{DBLP:conf/sigsoft/DavisCSL18,DBLP:conf/uss/StaicuP18}, path 
traversals~\cite{LiangThesis2018}, binding layer bugs~\cite{Brown2017}.

\paragraph{Publishing Model}
In order to publish a package, a developer needs to first create an account on 
the npm website. Once this prerequisite is met, adding a new package to the 
repository is as simple as running the ``npm publish'' command in a folder 
containing a package.json file. The user who first published 
the package is automatically added to the maintainers set and hence she can 
release future versions of that package. She can also decide to add additional 
npm users as maintainers. What is interesting to notice about this model is 
that it does not require a link to a public version control system, e.g., 
GitHub, hosting the 
code of the package. Nor does it require that persons who develop the code on 
such external repositories also have publishing rights on npm. This disconnect 
between the two platforms has led to 
confusion\footnote{\url{http://www.cs.tufts.edu/comp/116/archive/spring2018/etolhurst.pdf}}
in the past and to stealthy attacks 
that target npm accounts without changes to the versioning system.

\subsection{Threat Models}
\label{sec:tm}
The idiosyncratic security properties of npm, as described above, enable several scenarios for attacking users of npm packages.
The following discusses threat models that either correspond to attacks that have already occurred or that we consider to be possible in the future. 

\paragraph{Malicious Packages (TM-mal)}
Adversaries may publish packages containing malicious code on npm and hence trick other users into installing or depending on such packages.
In 2018, the eslint-scope incident mentioned earlier has been an example of this threat.
The package deployed its payload at installation time through an automatically executed post-installation script.
Other, perhaps more stealthy methods for hiding the malicious behavior could be envisioned, such as downloading and executing payloads only at runtime under certain conditions.

Strongly related to malicious packages are packages that violate the user's privacy by sending usage data to third parties, e.g., 
insight\footnote{\url{https://www.npmjs.com/package/insight}} or
analytics-node\footnote{\url{https://www.npmjs.com/package/analytics-node}}. 
While these libraries are legitimate under specific conditions, some users may not want to be tracked in this way.
Even though the creators of these packages clearly document the tracking functionality, transitive dependents may not be aware that one of their dependencies deploys tracking code.

\paragraph{Exploiting Unmaintained Legacy Code (TM-leg)}
As with any larger code base, npm contains vulnerable code, some of which is 
documented in public vulnerability databases such as npm security 
advisories\footnote{\url{https://www.npmjs.com/advisories}} or 
Snyk vulnerability DB\footnote{\url{https://snyk.io/vuln/?type=npm}}.
As long as a vulnerable package remains unfixed, an attacker can exploit it 
in applications that transitively depend on the vulnerable code.
Because packages may become abandoned due to developers 
inactivity~\cite{DBLP:journals/isse/ConstantinouM17} 
and because npm does not offer a forking mechanism, some packages may never be 
fixed.
Even worse, the common practice of locking dependencies may prevent applications from using fixed versions even when they are available.

\paragraph{Package Takeover (TM-pkg)}
An adversary may convince the current maintainers of a package to add her as a 
maintainer.
For example, in the recent \emph{event-stream} incident\footnote{\url{https://github.com/dominictarr/event-stream/issues/116}}, the attacker employed social engineering to obtain publishing rights on the target package.
The attacker then removed the original maintainer and hence became the sole owner of the package.
A variant of this attack is when an attacker injects code into the source 
base of the target package.
For example, such code injection may happen through a pull request, via compromised development tools, or even due to the fact that the attacker has commit rights on the repository of the package, but not npm publishing rights.
Once vulnerable or malicious code is injected, the legitimate maintainer would publish the package on npm, unaware of its security problems.
Another takeover-like attack is typosquatting, where an adversary publishes malicious code under a package name similar to the name of a legitimate, popular package.
Whenever a user accidentally mistypes a package name during installation, or a 
developer mistypes the name of a package to depend on, the malicious code will 
be installed.
Previous work shows that typosquatting attacks are easy to deploy and effective in practice~\cite{tschacher2016typosquatting}.

\paragraph{Account Takeover (TM-acc)}
The security of a package depends on the security of its maintainer accounts.
An attacker may compromise the credentials of a maintainer to deploy insecure code under the maintainer's name.
At least one recent incident (eslint-scope) is based on account takeover.
While we are not aware of how the account was hijacked in this case, there are various paths toward account takeover, e.g., weak passwords, social engineering, reuse of compromised passwords, and data breaches on npm. 


\paragraph{Collusion Attack (TM-coll)}
The above scenarios all assume a single point of failure.
In addition, the npm ecosystem may get attacked via multiple instances of the above threats.
Such a collusion attack may happen when multiple maintainers decide to conspire and to cause intentional harm, or when multiple packages or maintainers are taken over by an attacker.
\section{Methodology}
\label{sec:methodology}
To analyze how realistic  the above threats are, we systematically study 
package dependencies, maintainers, and known security vulnerabilities in npm.
The following explains the data and metrics we use for this study.

\subsection{Data Used for the Study}

\paragraph{Packages and Their Dependencies}

To understand the impact of security problems across the ecosystem, we analyze the dependencies between packages and their evolution.

\begin{definition}
Let $t$ be a specific point in time, $P_t$ be a set of npm package names, and 
$E_t 
= \{(p_i,p_j) | p_i \neq p_j \in 
P_t\}$ a set of directed edges between packages, where $p_i$ has a regular 
dependency on $p_j$. We call $G_t = (P_t,E_t)$ the {\bfseries npm dependency 
graph} at a given time $t$.
\end{definition}

We denote the universe of all packages ever published on npm with $\mathcal{P}$.
By aggregating the meta information about packages, we can easily construct the 
dependency graph without the need to download or install every package. Npm 
offers an API endpoint for downloading this metadata for all the releases of 
all packages ever published. In total we consider \nbPackageCount\ 
nodes and \nbEdgeCount\ edges.

To analyze the evolution of packages we gather data about all their releases.
As a convention, for any time interval $t$, such as years or months, we denote with $t$ the snapshot at the beginning of that time interval.
For example, $G_{2015}$ refers to the dependency graph at the beginning of the 
year 2015. In total we analyze \nbVersions\ releases, therefore an 
average of almost eight versions per package. Our observation period ends in 
April 2018.

\paragraph{Maintainers}

Every package has one or more developers responsible for publishing updates to the package.
\begin{definition}
For every $p \in P_t$, the set of {\bfseries maintainers $M(p)$} contains 
all users that have publishing rights for $p$. 
\end{definition}

Note that a specific user may appear as the maintainer of multiple packages and 
that the union of all maintainers in the ecosystem is denoted with $\mathcal{M} 
$.

\paragraph{Vulnerabilities}
The npm community issues advisories or public reports about vulnerabilities in 
specific npm packages. These advisories specify if there is a patch available 
and which releases of the package are affected by the vulnerability.

\begin{definition}
We say that a given package $p \in \mathcal{P}$ is 
{\bfseries vulnerable at a moment $t$} if there exists a public 
advisory for that package and if no patch was released for the described 
vulnerability at an earlier moment $t'<t$.
\end{definition}

We denote the set of vulnerable packages with $\mathcal{V} \subset \mathcal{P}$.
In total, we consider \nbCVEs\ advisories affecting 600 packages. We extract 
the 
data from the publicly available npm 
advisories\footnote{\url{https://www.npmjs.com/advisories}}.

\subsection{Metrics}
\label{sec:metrics}
We introduce a set of metrics for studying the risk of attacks on the npm 
ecosystem.

\paragraph{Packages and Their Dependencies}

The following measures the influence of a given package on other packages in the 
ecosystem.

\begin{definition}
For every $p \in P_t$, the {\bfseries package reach} $\mathrm{PR}(p)$ 
represents the set of all the packages that have a transitive dependency on $p$ 
in $G_t$.
\end{definition}

Note that the package itself is not included in this set. The reach 
$\mathrm{PR}(p)$ contains names of packages in the ecosystem. Therefore, the 
size of the set is bounded by the following values
$0\leq|\mathrm{PR}(p)| < |P_t|$.

Since $|\mathrm{PR}(p)|$ does not account for the ecosystem changes, the 
metric may grow 
simply because the ecosystem grows.
To address this, we also consider the average package reach:
\begin{equation}
\overline{\mathrm{PR_t}} = \frac{\sum_{\forall p \in P_t} | \mathrm{PR}(p) 
|}{|P_t|}
\end{equation}
Using the bounds discussed before for $\mathrm{PR}(p)$, we can calculate the 
ones for its average $0 \leq \overline{\mathrm{PR}}_t < |P_t|$. The upper 
limit is obtained for a fully connected graph in which all packages can reach 
all the other packages and hence $|PR(p)| = |P_t| - 1, \forall p$.
If $\overline{\mathrm{PR}}_t$ grows monotonously, we say that the 
ecosystem is getting more dense, and hence the average package influences an increasingly large number of packages. 

The inverse of package reach is a metric to quantify how many packages are 
implicitly trusted when installing a particular package.

\begin{definition}
For every $p \in P_t$, the set of {\bfseries implicitly trusted packages} 
$\mathrm{ITP}(p)$ contains all the packages $p_i$ for which \mbox{$p \in 
\mathrm{PR}(p_i)$}.\end{definition}

Similarly to the previous case, we also consider the size of the set 
$|\mathrm{ITP}(p)|$ and  the average number of 
implicitly trusted package $\overline{\mathrm{ITP}}_t$, having the same bounds 
as their package reach counterpart.

Even though the average 
metrics $\overline{\mathrm{ITP}}_t$ and $\overline{\mathrm{PR}}_t$ are 
equivalent for a given graph, the  distinction between their non-averaged 
counterparts is very important from a security point of view.
To see why, consider the example in Figure~\ref{fig:difftops}.
The average $\overline{PR}=\overline{ITP}$ is 5/6 = 0.83 both on the right and on the left.
However, on the left, a popular package $p1$ is dependent upon by many 
others. Hence, the package reach of $p1$ is five, and the number of implicitly 
trusted packages is one for each of the other packages.
On the right, though, the number of implicitly trusted packages for $p4$ is three, as users of $p4$ implicitly trust packages $p1$, $p2$, and $p3$.

\begin{figure}
\centering
\begin{subfigure}[b]{0.2\textwidth}  
\centering
\begin{tikzpicture}[->,>=stealth',shorten >=1pt,auto,node distance=1cm,
                    thick,main 
                    node/.style={circle,draw,font=\sffamily\footnotesize}]

  \node[main node] (1) {p1};
  \node[main node] (hidden) [below of=1, draw=none] {};  
  \node[main node] (3) [below left of=hidden] {p3};
   \node[main node] (2) [left of=3] {p2};
  \node[main node] (4) [right of=3] {p4};
  \node[main node] (5) [right of=4] {p5};
  \node[main node] (6) [above of=5] {p6};

  \path[every node/.style={font=\sffamily\footnotesize}]
    (2) edge node [] {} (1)
    (3) edge node [] {} (1)
    (4) edge node [] {} (1)
    (5) edge node [] {} (1)
    (6) edge node [] {} (1);

\end{tikzpicture}
\caption{Wide distribution of trust: $max(\mathrm{PR})=5, max(\mathrm{ITP}) = 
1$}
\end{subfigure}
\hspace{0.3cm}
\begin{subfigure}[b]{0.24\textwidth}  
\centering
\begin{tikzpicture}[->,>=stealth',shorten >=1pt,auto,node distance=1.2cm,
                    thick,main 
                    node/.style={circle,draw,font=\sffamily\footnotesize}]

  \node[main node] (1) {p1};
  \node[main node] (2) [below left of=1] {p2};
  \node[main node] (3) [below right of=1] {p3};
  \node[main node] (4) [below right of=2] {p4};
  \node[main node] (5) [right of=1] {p5};
  \node[main node] (6) [left of=1] {p6};

  \path[every node/.style={font=\sffamily\footnotesize}]
    (2) edge node [] {} (1)
    (3) edge node [] {} (1)
    (4) edge node [] {} (2)
    (4) edge node [] {} (3);

\end{tikzpicture}

\caption{Narrow distribution of trust: $max(\mathrm{PR})=3, 
max(\mathrm{ITP}) = 3$}
\end{subfigure}
\caption{Dependency graphs with different maximum package reaches ($\mathrm{PR}$) and different maximum numbers of trusted packages ($\mathrm{ITP}$).}
\label{fig:difftops}
\end{figure}

\paragraph{Maintainers}
The number of implicitly trusted packages or the package reach are important 
metrics for reasoning about TM-pkg, but not about TM-acc. That is because 
users may decide to split their functionality across multiple 
micropackages for which they are the sole maintainers. To put it differently, a 
large attack surface for  TM-pkg does not imply one for TM-acc.

Therefore, we define maintainer reach $\mathrm{MR}_t(m)$ and 
implicitly trusted maintainers $\mathrm{ITM}_t(p)$ for showing the influence of 
maintainers.

\begin{definition}
Let $m$ be an npm maintainer. The {\bfseries maintainer reach} $\mathrm{MR}(m)$ 
is the combined reach of all the maintainer's packages, $\mathrm{MR}(m) = \cup_{m \in 
M(p)} \mathrm{PR}(p)$
\end{definition}

\begin{definition}
For every $p \in P_t$, the set of {\bfseries implicitly trusted maintainers} 
$\mathrm{ITM}(p)$ contains all the maintainers that have publishing rights on 
at least one implicitly trusted package, $\mathrm{ITM}(p) = \cup_{p_i \in 
\mathrm{ITP}(p)}M(p_i)$.\end{definition}

The above metrics have the same bounds as their packages counterparts. Once 
again, the distinction between the package and the maintainer-level metrics is 
for shedding light on the security relevance of 
human actors in the ecosystem.

Furthermore, to approximate the maximum damage that colluding maintainers can 
incur on the 
ecosystem (TM-coll), we define an order in which the colluding 
maintainers are selected:

\begin{definition}
We call an ordered set of maintainers $L \subset \mathcal{M}$ a {\bfseries 
desirable 
collusion strategy} iff $\forall m_i \in L$ there is no $m_k \neq m_i$ for 
which \mbox{
$\cup_{j<i}\mathrm{MR}(m_j) \cup \mathrm{MR}(m_i) < \cup_{j<i}\mathrm{MR}(m_j) 
\cup \mathrm{MR}(m_k)$}.
\end{definition}

Therefore, the desirable collusion strategy is
a hill climbing algorithm in which at each step we choose the maintainer that 
provides the highest local increase in package reach at that point. We 
note that the problem of finding the set of $n$ maintainers that cover the most 
packages is an NP-hard problem called \emph{maximum coverage 
problem}. Hence, we believe 
that the proposed solution is a good enough approximation that shows how 
vulnerable the ecosystem is to a collusion attack, but that does not necessary 
yield the optimal solution.

\paragraph{Vulnerabilities}

For reasoning about TM-leg, we need to estimate how much of the ecosystem 
depends on vulnerable code:

\begin{definition}
Given all vulnerable packages $p_i \in \mathcal{V}_t$ at time $t$, we define 
the {\bfseries reach of vulnerable code} at time $t$ as
\mbox{$\mathrm{VR}_t = \cup_{p_i \in \mathcal{V}_t}PR(p_i)$}.
\end{definition}

Of course the actual reach of vulnerable code can not be fully calculated since 
it would rely on \emph{all} vulnerabilities present in npm modules, not only on 
the published ones. However, since in TM-leg we are interested in publicly 
known vulnerabilities, we define our metric according to this scenario. 
In these conditions, the speed at which vulnerabilities are reported is an 
important factor to consider: 

\begin{definition}
Given all vulnerable packages $p_i \in \mathcal{V}_t$ at time $t$, we define 
the {\bfseries vulnerability reporting rate} $\mathrm{VRR}_t$ at time $t$ as 
$\mathrm{VRR}_t = \frac{|\mathcal{V}_t|}{|P_t|}$.
\end{definition}

\section{Results}

We start by reporting the results on the nature of package level 
dependencies and their evolution over time (corresponding to TM-mal and 
TM-pkg). We then discuss the influence that maintainers have in the ecosystem 
(related to TM-acc and TM-coll). Finally, we explore the dangers of depending 
on unpatched security vulnerabilities (addressing TM-leg).

\subsection{Dependencies in the Ecosystem}
\label{subsec:deps}
\begin{figure}
	\includegraphics[width=\linewidth]{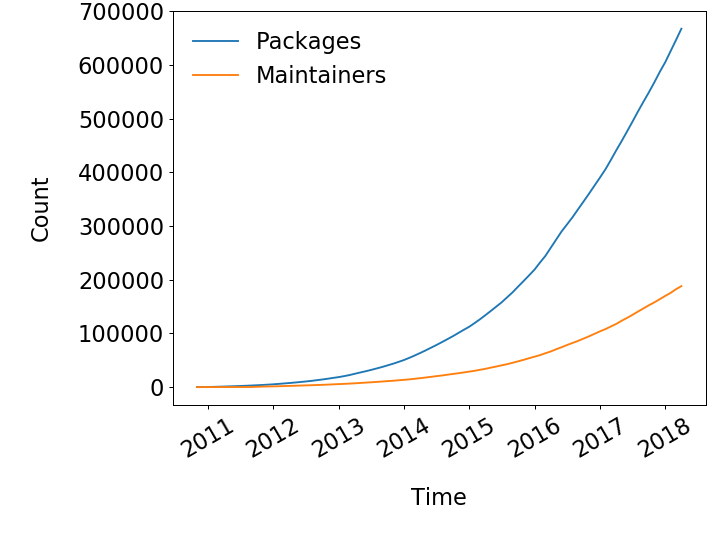}
	\caption{Evolution of number of packages and maintainers.}
	\label{fig:packagesMaintainersOverTime}
\end{figure}

To set the stage for a thorough analysis of security risks entailed by the structure of the npm ecosystem, we start with a general analysis of npm and its evolution.
Since its inception in 2010, the npm ecosystem has grown from a small collection of packages maintained by a few people to the world's largest software ecosystem.
Figure~\ref{fig:packagesMaintainersOverTime} shows the evolution of the number of packages available on npm and the number of maintainers responsible for these packages.
Both numbers have been increasing super-linearly over the past eight years.
At the end of our measurement range, there is a total of \nbPackageCount\ packages, a number likely to exceed one million in the near future.
These packages are taken care of by a total of \nbMaintainerCount\ maintainers. 
The ratio of packages to maintainers is stable across our observation period (ranging between 2.81 and 3.51).

In many ways, this growth is good news for the JavaScript community, as it increases the code available for reuse.
However, the availability of many packages may also cause developers to depend 
on more and more third-party code, which increases the attack surface for 
TM-pkg by giving individual packages the ability to impact the security of many 
other packages.
The following analyzes how the direct and transitive dependencies of packages are evolving over time (Section~\ref{sec:directAndTransitiveDependencies}) and how many other packages individual packages reach via dependencies (Section~\ref{sec:packageReach}).

\subsubsection{Direct and Transitive Dependencies}
\label{sec:directAndTransitiveDependencies}

\begin{figure}
	\includegraphics[width=\linewidth]{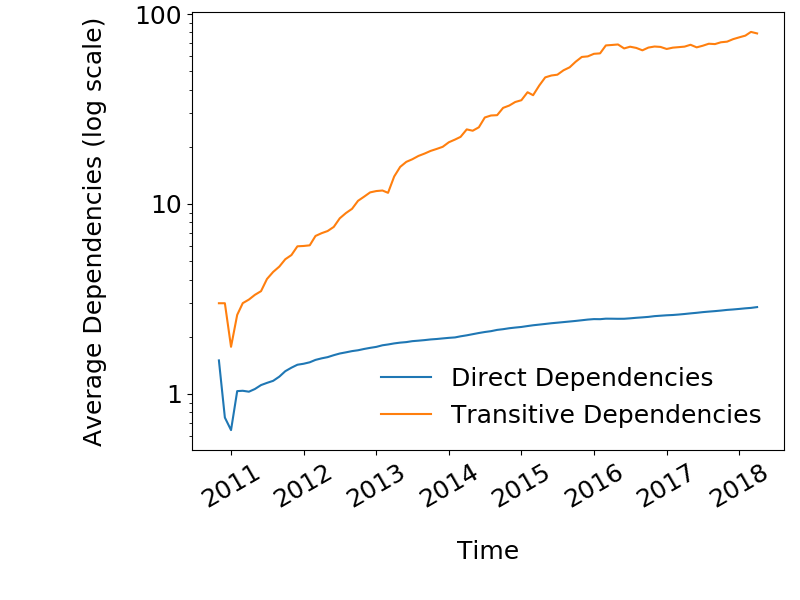}
	\caption{Evolution of direct package dependencies and its impact on 
	transitive dependencies. Note the logarithmic scale on the y-axis.}
	\label{fig:directAndTransitiveDependencies}
\end{figure}

Figure~\ref{fig:directAndTransitiveDependencies} shows how many other packages an average npm package depends on directly and transitively.
The number of direct dependencies has been increasing slightly from 1.3 in 2011 
to 2.8 in 2018, which is perhaps unsurprising given the availability of an 
increasing code base to reuse.
The less obvious observation is that a small, linear increase in direct dependencies leads to a significant, super-linear increase in transitive dependencies.
As shown by the upper line in Figure~\ref{fig:directAndTransitiveDependencies}, the number of transitive dependencies of an average package has increased to a staggering 80 in 2018 (note the logarithmic scale).

From a security perspective, it is important to note that each directly or transitively depended on package becomes part of the implicitly trusted code base.
When installing a package, each depended upon package runs its post-installation scripts on the user's machine -- code executed with the user's operating system-level permissions.
When using the package, calls into third-party modules may execute any of the code shipped with the depended upon packages.

\begin{resultbox}
	When installing an average npm package, a user implicitly trusts around 80 other packages due to transitive dependencies.
\end{resultbox}

One can observe in Figure~\ref{fig:directAndTransitiveDependencies} a chilling 
effect on the number of dependencies around the year 2016 which will become 
more apparent in the following graphs. Decan et 
al.~\cite{DBLP:journals/corr/abs-1710-04936} hypothesize that this effect 
is due to the left-pad incident. In order to confirm that this is not simply 
due to removal of more than a hundred packages belonging to the left-pad's 
owner, we remove all the packages owned by this maintainer. We see no 
significant difference for the trend in 
Figure~\ref{fig:directAndTransitiveDependencies} when removing these packages, 
hence we conclude that indeed there is a significant change in the structure of 
transitive dependencies in the ecosystem around 2016.

\subsubsection{Package Reach}
\label{sec:packageReach}

\begin{figure}
	\includegraphics[width=\linewidth]{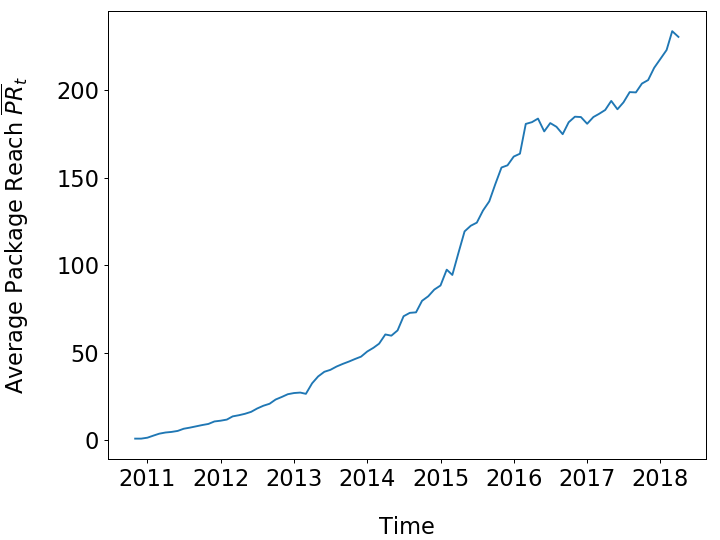}
	
    \includegraphics[width=\linewidth]{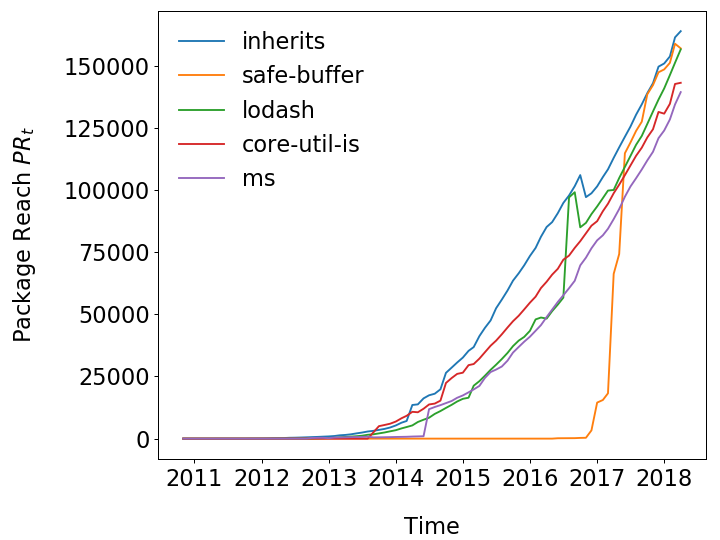}
	\caption{Evolution of package reach for an average package (top) and the 
	top-5 packages (bottom).}
	\label{fig:reach}
\end{figure}

The above analysis focuses on depended upon packages.
We now study the inverse phenomenon: packages impacted by individual packages, i.e., package reach as defined in Section~\ref{sec:methodology}.
Figure~\ref{fig:reach} shows how many other packages a single package reaches via direct or indirect dependencies.
The graph at the top is for an average package, showing that it impacts about \nbAvgPackageReachLatest\ other packages in 2018, a number that has been growing since the creation of npm.
The graph at the bottom shows the package reach of the top-5 packages (top in terms of their package reach, as of 2018).
In 2018, these packages each reach between \nbPackageReachTopMin\ and \nbPackageReachTopMax\ other packages, making them an extremely attractive target for attackers.

\begin{figure}
	\includegraphics[width=\linewidth]{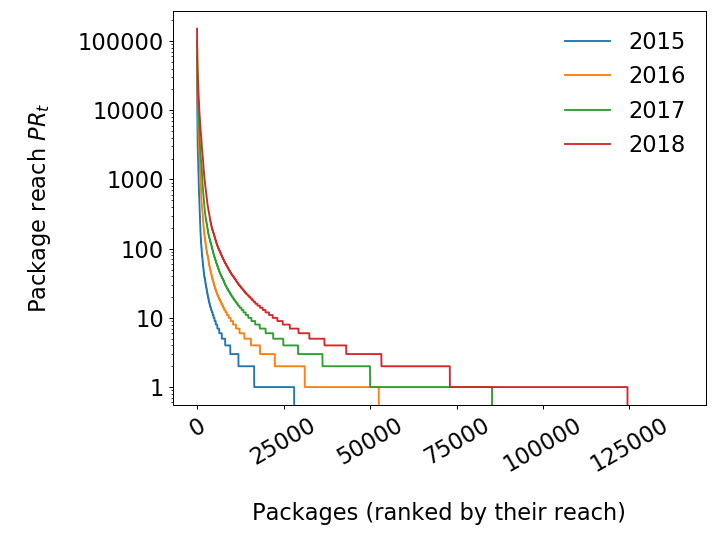}
	\caption{Distribution of package reach by individual packages, and how it 
	changes over time. Note the log scale on the vertical axis.}
	\label{fig:pkgReachYearEvolution}
\end{figure}

To better understand how the reach of packages evolves over time, Figure~\ref{fig:pkgReachYearEvolution} shows the distribution of reached packages for multiple years.
For example, the red line shows that in 2018, about 24,500 packages have 
reached 
at least 10 other packages, whereas only about 9,500 packages were so 
influential in 2015.
Overall, the figure shows that more and more packages are reaching a significant number of other packages, increasing the attractiveness of attacks that rely on dependencies.

\begin{resultbox}
	Some highly popular packages reach more than 100,000 other packages, making them a prime target for attacks.
	This problem has been aggravating over the past few years.
\end{resultbox}

The high reach of a package amplifies the effect of both vulnerabilities 
(TM-leg) and of malicious code (TM-mal).
As an example for the latter, consider the event-stream incident discussed 
when introducing TM-acc in Section~\ref{sec:tm}.
By computing event-stream's reach and comparing it with other packages, we see 
that this package is just one of many possible targets.
As of April 1, 2018 (the end of our measurement period), event-stream has a 
reach of 5,466.
That is, the targeted package is relatively popular, but still far from being 
the top-most attractive package to compromise.
In fact, 1,165 other packages have a greater or equal reach than event-stream.

\begin{resultbox}
Variants of the event-stream attack could easily be repeated with other packages.
\end{resultbox}

In order to perform a similar analysis for the eslint-scope security incident, 
we need to use a slightly modified version of package reach. This attack 
targeted a development tool, namely 
eslint, hence, to fully estimate the attack surface we need to 
consider dev dependencies in our definition of reach. We do not normally 
consider this type of
dependencies in our measurements because they are not automatically installed 
with a package, unlike regular dependencies. They are instead 
used only by the developers of the 
packages. Therefore the modified version of package reach considers both 
transitive regular dependencies and direct dev dependencies. 

We observe that eslint-scope has a modified reach of more than 100,000 packages 
at the last observation point in the data set. However, there are 347 other 
packages that have a higher reach, showing that even more serious attacks may 
occur in the future.

\begin{resultbox}
	The attack on eslint-scope has targeted a package with
	an influence  not larger than that of hundreds of other packages.
	It is likely that similar, or perhaps even worse, attacks will happen and 
	succeed in the future.
\end{resultbox}

\subsection{Analysis of Maintainers}
\label{subsec:maintainers}

We remind the reader that there is a significant difference between npm 
maintainers and repository contributors, as discussed in Section~\ref{sec:part}.
Even though contributors also have a lot of control over the code that will 
eventually end up in an npm package, they can not release a new version on npm, 
only the maintainers have this capability. Hence, the discussion that follows,  
about the security risks associated with maintainers, should be considered a 
lower bound for the overall attack surface.

Attacks corresponding to TM-acc in which maintainers are targeted are not 
purely hypothetical as the infamous eslint-scope
incident discussed earlier shows. In this attack, a malicious actor hijacked 
the account of an influential maintainer and then published a version of 
eslint-scope containing malicious code. This incident is a warning for how 
vulnerable the ecosystem is to targeted attacks and how maintainers influence 
can be used to deploy malware at scale. We further discuss the relation between 
packages and maintainers. 

\subsubsection{Packages per Maintainer}
Even though the ecosystem grows super-linearly as discussed in 
Section~\ref{subsec:deps}, 
one would expect that this is caused mainly by new developers joining the 
ecosystem. However, we observe that the number of packages per 
maintainer also grows suggesting that the current members of the platform are 
actively publishing new packages. The average number of packages controlled by 
a maintainer raises from 2.5 in 2012 to 3.5 in 2013 and almost 4.5 in 2018. 
Conversely, there are on average 1.35 maintainers in the lifetime of a package. 
The top 5,000 most popular packages have an average number of 2.83 maintainers. 
This is not unexpected, 
since multiple people are involved in developing the most popular packages, 
while for the majority of new packages there is only one developer.

\begin{figure}
	\includegraphics[width=\linewidth]{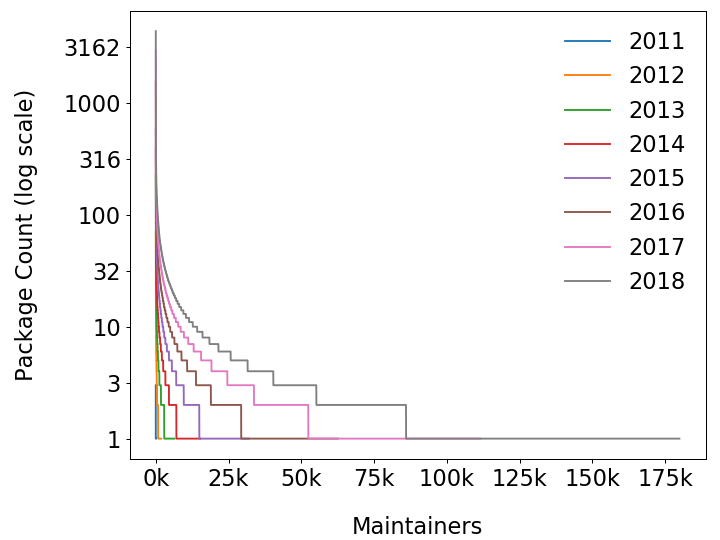}
	\caption{Evolution of maintainers sorted by package count per year.}
	\label{fig:evolutionMaintainers}
\end{figure}

Next, we study in more detail the evolution of the number of packages
a maintainer controls. Figure~\ref{fig:evolutionMaintainers} shows the 
maintainer package count plotted versus the number of maintainers having such a 
package count. Every line represents a year. The scale is logarithmic to base 
10. It shows that the majority of maintainers maintain few packages, yet some 
maintainers maintain over 100 packages. Over the years, the 
package count for the maintainers increased consistently. In 2015, only 
slightly more than 25,000 maintainers maintained more than one package, whereas 
this number has more than tripled by 2018.

\begin{figure}
	\includegraphics[width=\linewidth]{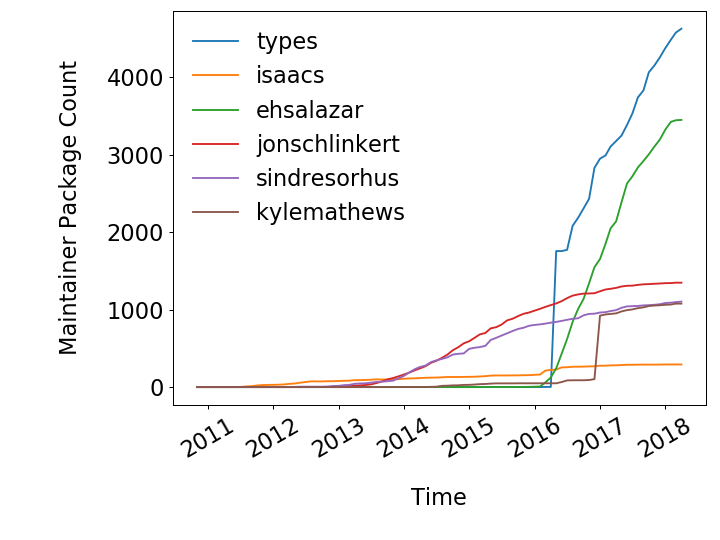}
	\caption{Evolution of package count for six popular maintainers.}
	\label{fig:packageCountSixMaint}
\end{figure}

We further analyze five different maintainers in top 20 according to number of 
packages and plot the evolution of 
their package count over the years in Figure~\ref{fig:packageCountSixMaint}. 
\emph{types} is the largest maintainer of 
type definitions for TypeScript, most likely a username shared by multiple 
developers at Microsoft, \emph{ehsalazar} maintains many security 
placeholder packages, \emph{jonschlinkert} and \emph{sindresorhus} are
maintaining many micropackages and \emph{isaacs} is the npm founder. From 
Figure~\ref{fig:packageCountSixMaint} we can see that for two of these 
maintainers the increase is superlinear or even near exponential: \emph{types} 
and \emph{kylemathews} have sudden spikes where they added many packages in a 
short time. We explain this by the tremendous increase in popularity for 
TypeScript in the recent years and by the community effort to prevent 
typosquatting attacks by reserving multiple placeholder. The graph of the other 
maintainers is more linear, but surprisingly it shows a continuous growth for 
all the six maintainers.

\begin{resultbox}
 The number of packages that both the influential and the average 
 maintainers control increased continuously over the years.
\end{resultbox}

\subsubsection{Implicitly Trusted Maintainers}
One may argue that the fact that maintainers publish new packages is a sign of 
a healthy ecosystem and that it only mimics its overall growth. However, we 
show that while that may be true, we also see an increase in the general 
influence of maintainers. That is, on average every package tends to 
transitively rely on more and more maintainers over time.

\begin{figure}
	\includegraphics[width=\linewidth]{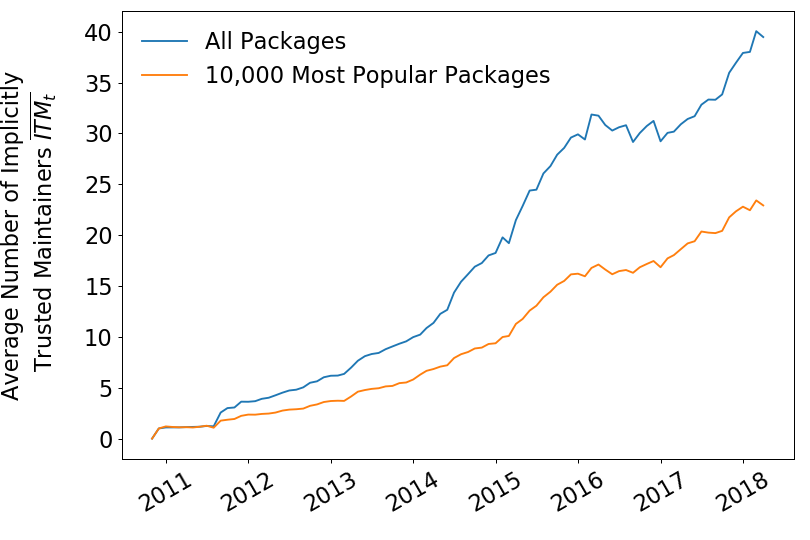}
	\caption{Evolution of average number of implicitly trusted maintainers over 
	years in all packages and in the most popular ones.}
	\label{fig:maintainersCostEvolution}
\end{figure}

In Figure~\ref{fig:maintainersCostEvolution} we show the evolution of 
$\overline{ITM_t}$, the average number of implicitly trusted maintainers. As 
can be seen, $\overline{ITM_t}$
almost doubled in the last three years for the average npm package, despite the 
plateau of the curve reached in 2016 which we again speculate it is caused by 
the left-pad incident. This is a worrisome development since compromising any 
of the maintainer accounts a package trusts may seriously impact the security 
of that package, as discussed in TM-acc. The positive aspect of 
the data in Figure~\ref{fig:maintainersCostEvolution} is that the growth in 
the number of implicitly trusted maintainers seems to be less steep for the top 
10,000 packages compared to 
the whole ecosystem. We hypothesize that the developers of popular packages are 
aware of this problem and actively try to limit the $\overline{ITM_t}$. 
However, a value over 20 for the average popular package is still high enough 
to be problematic.

\begin{resultbox}
The average npm package transitively relies on code published by 40 
maintainers. Popular packages rely on only 20.
\end{resultbox}

\begin{figure}
	\includegraphics[width=0.95\linewidth]{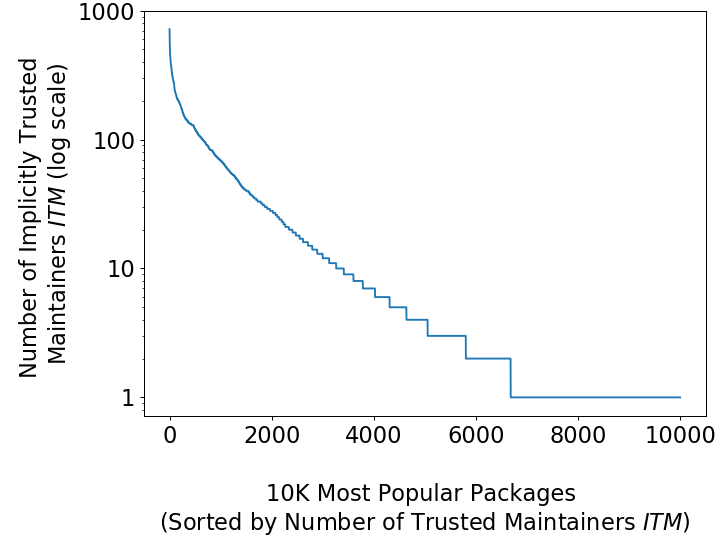}
	\caption{Number of implicitly trusted maintainers for top 10,000 most 
	popular packages.}
	\label{fig:maintainersCost10k}
\end{figure}

When breaking the average $\overline{ITM_t}$ discussed earlier into individual 
points in Figure~\ref{fig:maintainersCost10k},
one can observe that the majority of these packages can be influenced by more 
than one maintainer. This is surprising since most of the popular packages 
are micropackages such as "inherits" or "left-pad" or libraries with no 
dependencies like "moment" or "lodash". 
However, only around 30\% of 
these top packages have a maintainer cost higher than 10. Out of these, though, 
there are 643 packages influenced by more than a hundred maintainers. 

\begin{resultbox}
More than 600 highly popular npm packages rely on code published by at least 
100 maintainers.
\end{resultbox}

\subsubsection{Maintainers Reach}

\begin{figure}
	\includegraphics[width=0.95\linewidth]{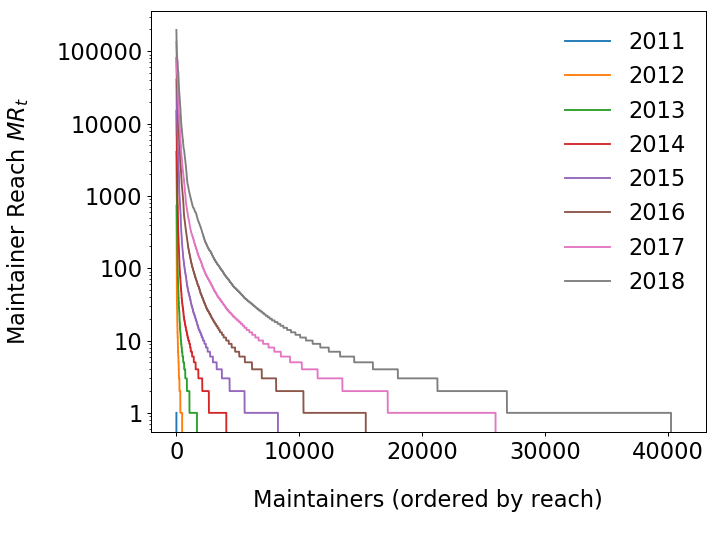}
	\caption{Distribution of maintainers reach in different years.}
	\label{fig:maintainersReachYear}
\end{figure}

\begin{figure}
	\includegraphics[width=\linewidth]{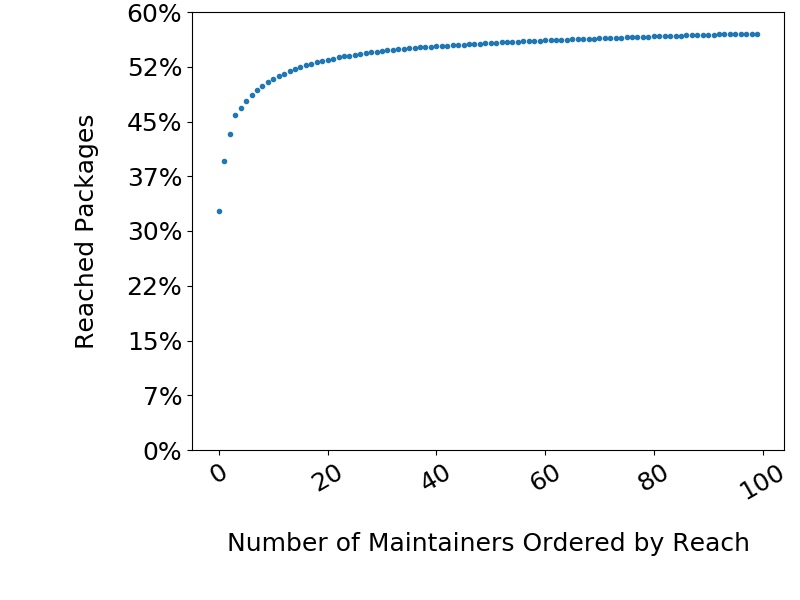}
	\caption{Combined reach of 100 influential maintainers.}
	\label{fig:reach100Maintainers}
\end{figure}

In Figure~\ref{fig:maintainersReachYear}, we plot the reach $MR_t$ of the 
maintainers in the npm ecosystem. The reach has increased over the 
years at all levels. For example, in 2015 there were 2,152 maintainers that 
could affect more than 10 packages, and this number increased to 4,041 in 
2016, 6,680 in 2017 and finally reaching an astonishingly high 10,534 in 2018. 
At the other end of the distribution, there were 59 maintainers that could 
affect more than 10,000 packages in 2015, 163 in 2016, 249 in 2017 and finally 
391 in 2018. The speed of growth for $MR_t$ is worrisome, 
showing that more and more developers have control over thousands of packages. 
If an attacker manages to compromise the account of any of the 391 most 
influential maintainers, the community will experience a serious security 
incident, reaching twice as many packages as in the event-stream attack.
\begin{resultbox}

	391 highly influential maintainers affect more than 10,000 packages, making 
	them prime targets for attacks.

	The problem has been aggravating over the past years.

\end{resultbox}
Finally, we look at the scenario in which multiple popular maintainers collude,
according to the desirable collusion strategy introduced in 
Section~\ref{sec:metrics}, to perform a large-scale attack on the 
ecosystem, 
i.e., TM-col. In 
Figure~\ref{fig:reach100Maintainers} we show that 20 maintainers can reach more 
than half of the ecosystem. Past that point every new maintainer joining 
does not increase significantly the attack's performance.

\subsection{Security Advisories Evolution}
\label{sec:adv-section}

\begin{figure}
	\includegraphics[width=0.95\linewidth]{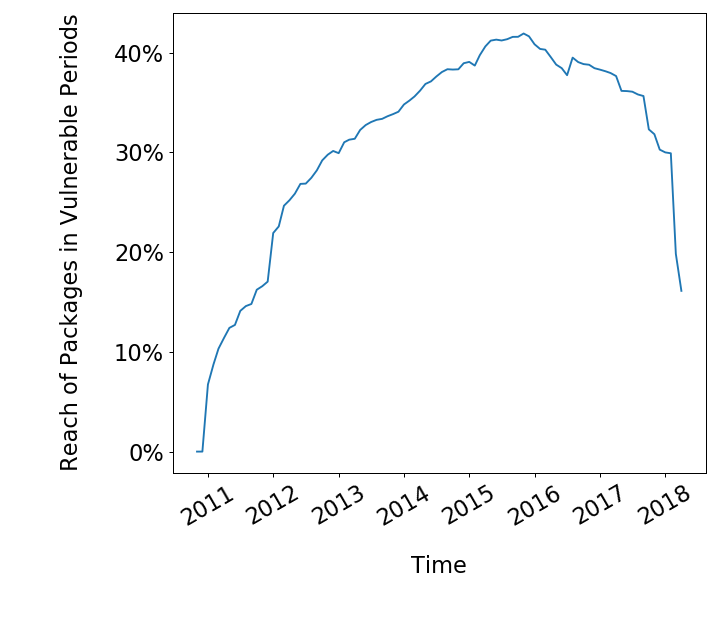}
	\caption{Total reach of packages for which there is at least one unpatched 
		advisory (vulnerability reach $\mathrm{VR}_t$).}
	\label{fig:cvesReachEvolution}
\end{figure}

\begin{figure}
	\includegraphics[width=0.95\linewidth]{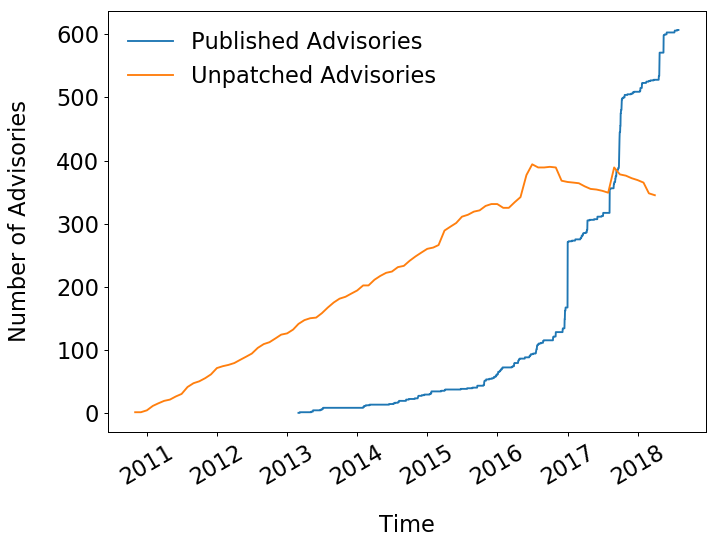}
	\caption{Evolution of the total and unpatched number of 
		advisories.}
	\label{fig:cvesEvolution}
\end{figure}

\begin{figure}
	\includegraphics[width=\linewidth]{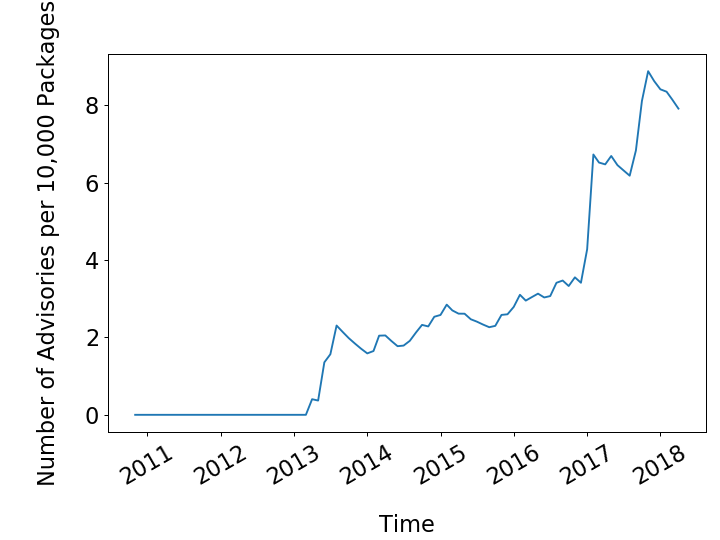}
	\caption{Evolution of $\mathrm{VRR}_t$, the rate of published 
		vulnerabilities per 10,000 
		packages.\cas{Should we define this rate per number of releases instead of 
			number of packages? If so, @Markus or @Cam: can you help me with 
			getting the data for this?}}
	\label{fig:cvesRateGrowth}
\end{figure}

Next, we study how often vulnerabilities are reported and fixed in the npm 
ecosystem (TM-leg). Figure~\ref{fig:cvesEvolution} 
shows the number of reported vulnerabilities in the lifetime of the ecosystem. 
The curve seems to resemble the evolution of number of 
packages presented in Figure~\ref{fig:packagesMaintainersOverTime}, with a 
steep increase in the last two years. To explore this relation further we plot 
in Figure~\ref{fig:cvesRateGrowth} the evolution of the number of advisories 
reported per 10,000 packages and we observe that it grows from two in 2013 to 
almost eight in 2018. This is a sign of a healthy security community that 
reports vulnerabilities at a very good pace, keeping up with the growth of the 
ecosystem. 

When analyzing the type of reported vulnerabilities in details, we 
observe that almost half of the advisories come from two 
large-scale campaigns and not a broader community effort: First, there are 
141 advisories published in January 2017 involving npm packages that download 
resources over HTTP, instead of HTTPs. Second, there are 120 directory 
traversal vulnerabilities reported as part of the research efforts of Liang 
Gong~\cite{LiangThesis2018}. Nevertheless, this shows the feasibility of 
large-scale vulnerability detection and reporting on npm.

Publishing an advisory helps raise awareness of a security problem in an npm 
package, but in order to keep the users secure, there needs to be a patch 
available for a given advisory. In Figure~\ref{fig:cvesEvolution}  we 
show the evolution of the number of unpatched security vulnerabilities in npm, 
as defined in Section~\ref{sec:methodology}. 
This trend is alarming, suggesting that two out of three advisories are still 
unpatched, leaving the users at risk. When manually inspecting some of the 
unpatched advisories we notice that a large percentage of unpatched 
vulnerabilities are actually advisories against malicious typosquatting 
packages for which no fix can be available. 

To better understand the real impact of the unpatched vulnerabilities we 
analyze how much of the ecosystem they impact, i.e., vulnerability reach as 
introduced in Section~\ref{sec:metrics}. To that end, 
we compute the reach of unpatched packages at every point in time in 
Figure~\ref{fig:cvesReachEvolution}. At a first sight, this data shows a much 
less grim picture than expected, suggesting that the reach of vulnerable 
packages is dropping over time. However, we notice that the effect of 
vulnerabilities tends to be retroactive. That is, a vulnerability published in 
2015 affects multiple versions of a package released prior to that date, hence 
influencing the data points corresponding to the years 2011-2014 in
Figure~\ref{fig:cvesReachEvolution}. Therefore, 
the vulnerabilities that will be reported in the next couple of years may 
correct for the downwards trend we see on the graph.
\cas{I think it's out of scope for this submission, but we said that it would 
be interesting to have a qualitative analysis of the unpatched vulnerabilities, 
i.e. split them in different classes: typosquatting, malware, abandoned, etc.}
Independent of the downwards trend, the fact that for the majority of the time 
the reach of vulnerable unpatched code is between 30\% and 40\% is alarming.

\begin{resultbox}
	Up to 40\% of all packages rely on code known to be vulnerable.
\end{resultbox}

\section{\CHANGED{Potential Mitigations}}
\label{sec:mitigations}

\CHANGED{The following section discusses ideas for mitigating some of the security threats in the npm ecosystem.
We here do not provide fully developed solutions, but instead outline ideas for future research, along with an initial assessment of their potential and challenges involved in implementing them.}

\subsection{Raising Developer Awareness}

\begin{figure}
	\includegraphics[width=0.95\linewidth]{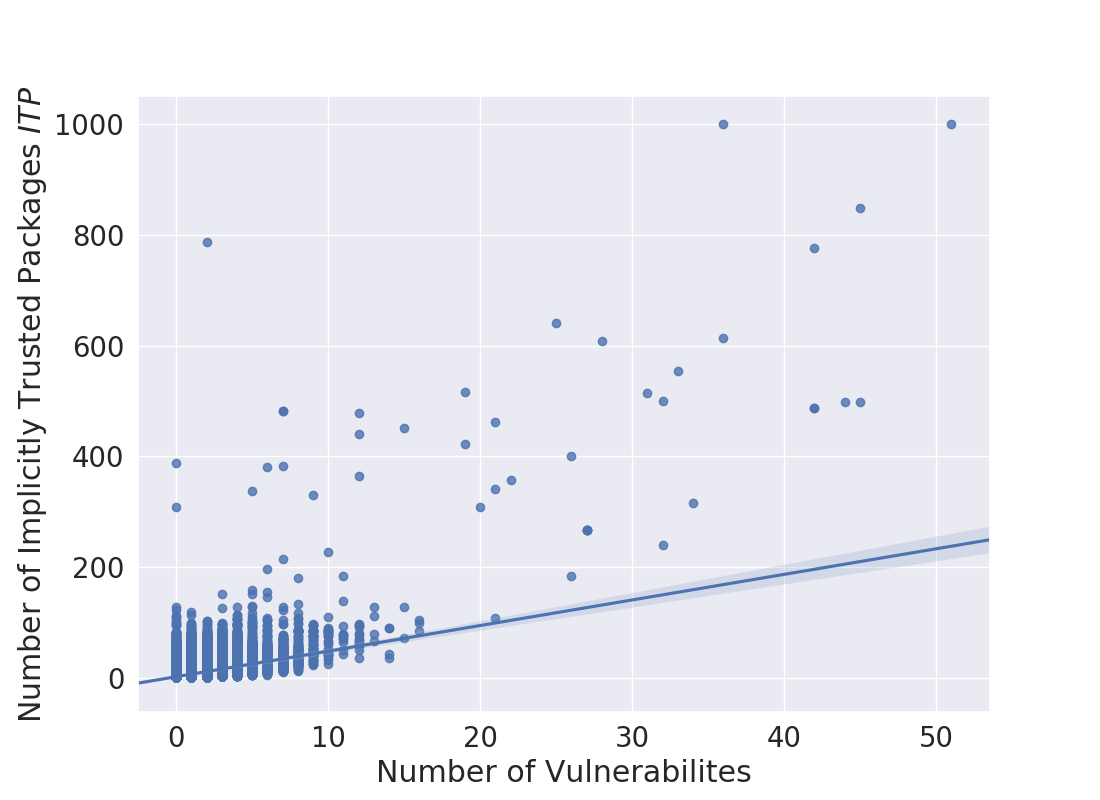}
	\caption{Correlation between number of vulnerabilities and number of 
		dependencies.}
	\label{fig:correl}
\end{figure}

One line of defense against the attacks described in this paper is to make developers who use third-party packages more aware of the risks entailed by depending on a particular package.
Currently, npm shows for each package the number of downloads, dependencies, dependents, and open issues in the associated repository.
However, the site does not show any information about the transitive dependencies or about the number of maintainers that may influence a package, i.e., our $\mathrm{ITP}$ and $\mathrm{ITM}$ metrics.
As initial evidence that including such metrics indeed predicts the risk of security issues, Figure~\ref{fig:correl} shows the number of implicitly trusted packages versus the number of vulnerabilities a package is affected by.
We find that the two values are correlated (Pearson correlation coefficient of 0.495), which is not totally unexpected since adding more dependencies increases the chance of depending on vulnerable code.
Showing such information, e.g., the $\mathrm{ITP}$ metric, could help developers make more informed decisions about which third-party packages to rely on.

\subsection{Warning about Vulnerable Packages}

To warn developers about unpatched vulnerabilities in their dependencies, the \code{npm audit} tool has been introduced.
It compares all directly depended upon packages against a database of known vulnerabilities, and warns a developer when depending upon a vulnerable version of a package.
While being a valuable step forward, the tool currently suffers from at least three limitations.
First, it only considers direct dependencies but ignores any vulnerabilities in transitive dependencies.
Second, the tool is limited to known vulnerabilities, and hence its effectiveness depends on how fast advisories are published.
Finally, this defense is insufficient against malware attacks.

%

\subsection{Code Vetting}

\begin{figure}
	\includegraphics[width=\linewidth]{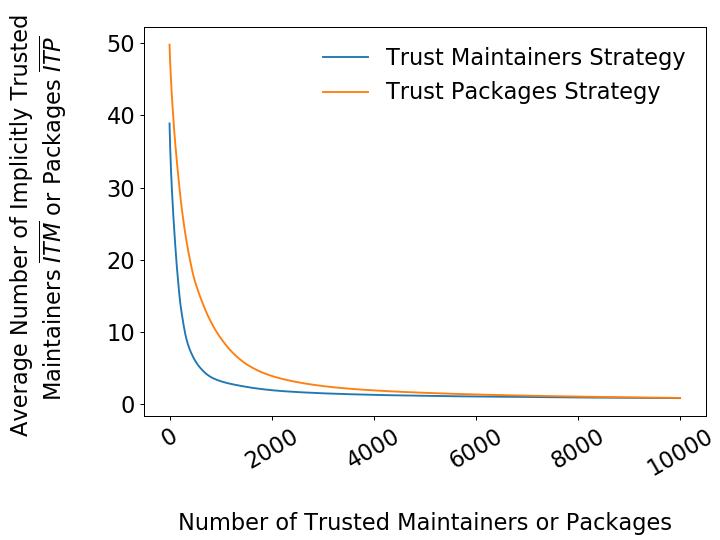}
	\caption{Decrease in average number of implicitly trusted maintainers and 
		packages as the set of trusted maintainers or packages increases.\todo{``Trust'' should be ``Trusted''}}
	\label{fig:trustedMaintainers}
\end{figure}

A proactive way of defending against both vulnerable and malicious code is code vetting.
Similar to other ecosystems, such as mobile app stores, whenever a new 
release of a vetted package is published, npm could analyze its code.
If and only if the analysis validates the new release, it is made available to users.
Since the vetting process may involve semi-automatic or even manual steps, we believe that it is realistic to assume 
that it will be deployed step by step in the ecosystem, starting with the most popular packages.
Figure~\ref{fig:trustedMaintainers} (orange curve) illustrates the effect that such code vetting could have on the ecosystem.
The figure shows how the average number of implicitly trusted packages, $\overline{\mathrm{ITP}}$, reduces with an increasing number of vetted and therefore trusted packages.
For example, vetting the most dependent upon 1,500 packages would reduce the $\overline{\mathrm{ITP}}$ ten fold, and vetting 4,000 packages would 
reduce it by a factor of 25.

\begin{figure}
	\includegraphics[width=0.95\linewidth]{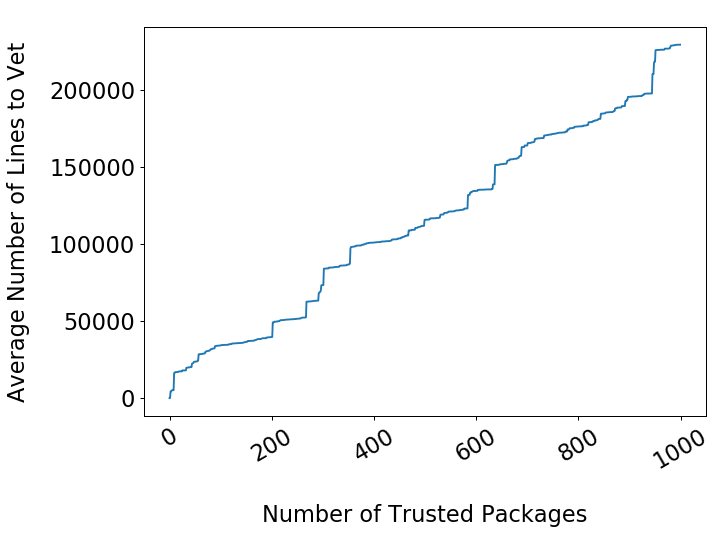}
	\caption{Number of lines of code that need to be vetted for achieving a 
		certain number of trusted packages.}
	\label{fig:churn}
\end{figure}

An obvious question is how to implement such large-scale code vetting, in particular, given that new versions of packages are released regularly.
To estimate the cost of vetting new releases, Figure~\ref{fig:churn} shows the average number of lines of code that are changed per release of a package, and would need to be vetted to maintain a specific number of trusted packages.
For example, vetting the changes made in a single new release of the top 400 most popular packages requires to analyze over 100,000 changed lines of code.
One way to scale code vetting to this amount of code could be automated code analysis tools.
Recently, there have been several efforts for improving the state of the art of security auditing for npm, both from academia, e.g.,
Synode~\cite{DBLP:conf/ndss/StaicuPL18}, 
BreakApp~\cite{DBLP:conf/ndss/VasilakisKRDDS18}, 
NodeSec~\cite{LiangThesis2018}, NoRegrets~\cite{DBLP:conf/ecoop/MezzettiMT18}, 
Node.cure~\cite{DBLP:conf/uss/DavisWL18}, and from industry practitioners, 
e.g., Semmle\footnote{\url{https://semmle.com/}},
r2c\footnote{\url{https://r2c.dev/}}, and
DeepScan\footnote{\url{https://deepscan.io/}}.
Orthogonal to automated code analysis tools, the npm community could establish crowd-sourced package vetting, e.g., in a hierarchically organized code distribution model similar to the Debian ecosystem.


Another challenge for code vetting is that npm packages, in contrast to apps in mobile app stores, are used across different platforms with different security models.
For example, XSS vulnerabilities are relevant only when a package is used on the client-side, whereas command injection via the \code{exec} API~\cite{DBLP:conf/ndss/StaicuPL18} is a concern only on the server-side.
A code vetting process could address this challenge by assigned platform-specific labels, e.g., ``vetted for client-side'' and ''vetted for server-side'', depending on which potential problems the vetting reveals.

\subsection{Training and Vetting Maintainers}

Another line of proactive defense could be to systematically train and vet highly influential maintainers.
For example, this process could validate the identity of maintainers, support maintainers in understanding basic security principles, and ensure that their accounts are protected by state-of-the-art techniques, such as two-factor authentication.
To assess the effect that such a process would have, we simulate how training and vetting a particular number of \emph{trusted maintainers} influences the average number of 
implicitly trusted maintainers, $\overline{\mathrm{ITM}}$.
The simulation assumes that the most influential maintainers are vetted first, and that once a maintainer is vetted she is ignored in the computation of the $\overline{\mathrm{ITM}}$.
The results of this simulation (Figure~\ref{fig:trustedMaintainers}) show a similar effect as for vetting packages: Because some maintainers are highly influential, vetting a relatively small number of maintainers can significantly reduce security risks.
For example, vetting around 140 maintainers cuts down the $\overline{\mathrm{ITM}}$ in half, and vetting around 600 could even reduce $\overline{\mathrm{ITM}}$ to less than five.
These results show that this mechanism scales reasonably well, but that 
hundreds of maintainers need to be vetted to bring the average number of 
implicitly trusted maintainers to a reasonable level.
Moreover, two-factor authentication has its own risks, e.g., when developers handle authentication tokens in an insecure way\footnote{\url{https://blog.npmjs.org/post/182015409750/automated-token-revocation-for-when-you}} or when attackers attempt to steal such tokens, as in the eslint-scope incident.

\section{Related Work}
In this section we discuss the closest related work contained mainly in two 
distinct research areas: JavaScript security and software 
ecosystem studies. While some of this work studies the npm ecosystem, to the 
best of our knowledge, we are the first to analyze in depth the role 
maintainers play in the ecosystem and the impact of different types of attacks, 
as well as the potential impact of vetting code.

\paragraph{Server-side JavaScript Security} 
There are many studies that investigate problems with dependency management for 
the JavaScript or other ecosystems. Abdalkareem et 
al.~\cite{DBLP:conf/sigsoft/AbdalkareemNWMS17} investigate reasons why 
developers would use trivial packages. 
They find that developers think that these packages are well implemented and 
tested and that they increase productivity 
as the developer does not need to implement such small features herself. 
Another empirical study on micropackages by Kula et 
al.~\cite{DBLP:journals/corr/abs-1709-04638} has similar results. They show 
that micropackages have long dependency chains, something we also discovered 
in some case studies of package reach. We also show that these packages have a 
high potential of being a target of an attack as they are dependent on by a lot 
of packages. 
Another previously studied topic is breaking changes introduced by 
dependencies. Bogart et al.~\cite{DBLP:conf/sigsoft/BogartKHT16} perform a case 
study interviewing developers about breaking changes in three different 
ecosystems. They 
find that npm's community values a fast approach to new releases compared to 
the other ecosystems. Developers of npm are more willing to adopt breaking 
changes to fight technical debt. Furthermore, they find that the semantic 
versioning rules are enforced more overtime than in the beginning. Similarly, 
Decan et 
al.~\cite{DBLP:conf/wcre/DecanMC17} analyze three package ecosystems, including 
npm, and evaluate whether dependency constraints and semantic 
versioning are effective measures for avoiding breaking changes. They find that 
both these measures are not perfect and that there is a need for better 
tooling. One such tool can be the testing  technique by Mezzetti et 
al.~\cite{DBLP:conf/ecoop/MezzettiMT18} which 
automatically detects whether an update of a package contains a breaking change 
in the API. With this method, they can identify type-related breaking changes 
between two versions. They identify 26 breaking changes in 167 updates of 
important npm packages. Pfretzschner et 
al.~\cite{DBLP:conf/IEEEares/PfretzschnerO17} describe four possible 
dependency-based attacks that exploit weaknesses such as global variables or 
monkeypatching in Node.js. They implement a detection of such attacks, but they 
do not find any real-world exploits. 
One way to mitigate these attacks is implemented by Vasilakis et 
al.~\cite{DBLP:conf/ndss/VasilakisKRDDS18} in
\textit{BreakApp}, a tool that creates automatic compartments for each 
dependency to enforce security policies. This increases security when using 
untrusted third-party packages.
Furthermore, third-party packages can have security vulnerabilities that can 
impact all the dependents. Davis et al.~\cite{DBLP:conf/sigsoft/DavisCSL18} and 
Staicu et al.~\cite{DBLP:conf/uss/StaicuP18} find denial of service 
vulnerabilities in regular expressions in the npm ecosystem. In another study, 
Staicu et al.~\cite{DBLP:conf/ndss/StaicuPL18} find several injection 
vulnerabilities due to the \textit{child\_process} module or the \textit{eval} 
function. 
Brown et al.~\cite{Brown2017} discuss bugs in the binding layers of both 
server-side and client-side JavaScript platforms, while  Wang et 
al.~\cite{DBLP:conf/kbse/WangDGGQYW17} analyze concurrency bugs in Node.js
Finally, Gong~\cite{LiangThesis2018} presents a dynamic analysis system for 
identifying 
vulnerable and malicious code in npm. He reports more than 300 previously 
unknown vulnerabilities, some of which are clearly visible on the figures in 
Section~\ref{sec:adv-section}.
Furthermore, there are studies that look at how frequent security 
vulnerabilities are in the npm ecosystem, how fast packages fix these and how 
fast dependent packages upgrade to a non-vulnerable version. Chatzidimitriou et 
al.~\cite{DBLP:conf/msr/Chatzidimitriou18} build an infrastructure to measure 
the quality of the npm ecosystem and to detect publicly disclosed 
vulnerabilities in package dependencies.
Decan et al.~\cite{DBLP:conf/msr/DecanMC18} perform a similar study but 
they investigate the evolution of vulnerabilities over time. They find that 
more than half of the dependent packages are still affected by a vulnerability 
after the fix is released. However, we show that the problem is even more 
serious because for more than half of the npm packages there is no available 
patch.

\paragraph{Client-Side (JavaScript) Security}
Client-side security is a vast and mature research area and it is 
out scope to extensively survey it here. Instead, we focus on those studies 
that analyze dependencies in client-side code. Nikiforakis et 
al.\cite{Nikiforakis2012} present a study of remote inclusion of JavaScript 
libraries in the most popular 10,000 websites. They show that an average 
website in their data set adds between 1.5 and 2 new dependencies per year. 
Similar to our work, they then discuss several threat models and attacks that 
can occur in this tightly connected ecosystem. Lauinger et 
al.~\cite{Lauinger2017} study the inclusion of libraries with known 
vulnerabilities in both popular and average 
websites. They show that 37\% of the websites in their data set include at 
least one vulnerable library. This number is suprisingly close to the reach we 
observe in npm for the vulnerable code. However, one should take both these 
results with a grain of salt since inclusion of vulnerable libraries does not 
necessary lead to a security problem if the library is used in a safe way. 
Libert et al.~\cite{DBLP:journals/corr/Libert15} perform a HTTP-level analysis 
of third-party resource inclusions, i.e., dependencies. They conclude that nine 
in ten websites leak data to third-parties and that six in ten spwan 
third-party cookies. 

\paragraph{Studies of Software Ecosystems} Software ecosystem 
research has been rapidly growing in the last year. 
Manikas~\cite{DBLP:journals/jss/Manikas16} surveys the related work and 
observes a maturing field at the intersection of multiple other research areas. 
Nevertheless, he identifies a set of challenges, for example, the 
problem of generalizing specific ecosystem research to other ecosystems or the 
lack of theories specific to software ecosystems. Serebrenik et 
al.~\cite{DBLP:conf/ecsa/SerebrenikM15} perform a meta-analysis of the 
difficult tasks in software ecosystem research and identify six types of 
challenges. For example, how to scale the analysis to the massive amount of 
data, how to research the quality and evolution of the ecosystem and how to 
dedicate more attention to comparative studies. 
Mens~\cite{DBLP:conf/icsm/Mens16} further looks at the 
socio-technical view on software maintenance and evolution. He argues that 
future research needs to study both the technical and the social dimension of 
the ecosystem. 
Our study follows this recommendation as it not only looks at 
the influence of a package on the npm ecosystem, but also at the influence of 
the maintainers. 
Several related work advocates metrics borrowed from other 
fields. For example, Lertwittayatrai et 
al.~\cite{DBLP:conf/apsec/Lertwittayatrai17} use network analysis techniques 
to study the topology of the JavaScript package ecosystem and to extract 
insights 
about dependencies and their relations.
Another study by Kabbedijk et 
al.~\cite{DBLP:conf/icsob/KabbedijkJ11} looks at the social aspect of the 
Ruby software ecosystem by identifying different roles maintainers have in the 
ecosystem, depending on the number of developers they cooperate with and on the 
popularity of their packages.
Overall, the research field is rising with a lot of studied software 
ecosystems in addition to the very popular ones such as JavaScript which is 
the focus of our study. 

\paragraph{Ecosystem Evolution}
Studying the evolution of an ecosystem shows how fast it grows and whether 
developers still contribute to it. Wittern et 
al.~\cite{DBLP:conf/msr/WitternSR16} study the whole JavaScript ecosystem,
including GitHub and npm until September 2015. They focus on dependencies, the 
popularity of packages and version numbering. They find that the ecosystem is 
steadily growing and exhibiting a similar effect to a power law distribution as 
only a quarter of packages is dependent upon. Comparing these numbers with our 
results, we see a continuous near-exponential growth in the number of released 
packages and that only 20\% of all packages are dependent upon. 
A similar study that includes the JavaScript ecosystem by Kikas et 
al.\cite{DBLP:conf/msr/KikasGDP17} collects data until May 2016 and focuses on 
the evolution of dependencies and the vulnerability of the dependency network. 
They confirm the same general growth as the previous study. Furthermore, they 
find packages that have a high impact with up to 30\% of other 
packages and applications affected.
Our study gives an update on these studies and additionally looks at the 
evolution of maintainers as they are a possible vulnerability in the 
ecosystem. 
The dependency network evolution was also studied for other ecosystems. Decan 
et al.~\cite{DBLP:journals/corr/abs-1710-04936} compare the evolution of seven 
different package managers focusing on the dependency network. Npm is the 
largest ecosystem in their comparison and they discover that dependencies are 
frequently used in all these ecosystems with similar connectedness between 
packages. Bloemen et al.~\cite{DBLP:conf/msr/BloemenAKO14a} look at software 
package dependencies of the Linux distribution \textit{Gentoo} where they use 
cluster analysis to explore different categories of software. German et 
al.~\cite{DBLP:conf/csmr/GermanAH13} study the dependency network of the 
\textit{R} language and the community around its
user-contributed packages. Bavota et al.~\cite{DBLP:conf/icsm/BavotaCPOP13a} 
analyze the large Apache ecosystem of Java libraries where they find that while 
the number of projects grows linearly, the number of dependencies between them 
grows exponentially. Comparing this to the npm ecosystem, we find the number of 
packages to grow super-linearly while the average number of dependencies 
between them grows linearly. 

\section{Conclusions}

We present a large-scale study of security threats resulting from the densely connected structure of npm packages and maintainers.
The overall conclusion is that npm is a small world with high risks.
It is ``small'' in the sense that packages are densely connected via dependencies.
The security risk are ``high'' in the sense that vulnerable or malicious code 
in a single package may affect thousands of others, and that a single 
misbehaving maintainer, e.g., due to a compromised account, may have a huge 
negative impact.
These findings show that recent security incidents in the npm ecosystem are 
likely to be the first signs of a larger problem, and not only unfortunate 
individual cases.
To mitigate the risks imposed by the current situation, we analyze the potential effectiveness of several mitigation strategies.
We find that trusted maintainers and a code vetting process for selected 
packages could significantly reduce current risks.

\medskip
\noindent
\textbf{Acknowledgments}
\begin{spacing}{0.9} \small
\noindent
This work was supported by the German Federal Ministry of Education and 
Research and by the Hessian Ministry of Science and the Arts within CRISP, by 
the German Research Foundation within the ConcSys and Perf4JS projects. The
authors would also like to thank the team at r2c for their
engineering support in obtaining the data for this work.
\end{spacing}

\bibliographystyle{plain}
\bibliography{references}

\end{document}